\documentclass[twocolumn]{IEEEtran}

\usepackage{algorithm}
\usepackage{lineno}
\usepackage{algorithmic}
\usepackage{algorithm}
\usepackage{array}
\usepackage{multicol}
\usepackage{amsfonts}

\usepackage{epsfig}
\usepackage{amsmath}
\usepackage{latexsym}
\usepackage{amssymb}
\usepackage{amscd}
\usepackage{graphicx}
\usepackage{amsmath}
\usepackage{amssymb}
\usepackage{bm}
\usepackage{amssymb}
\usepackage{amsmath}

\newcommand{\y}{\mathbf{y}}
\newcommand{\x}{\mathbf{x}}
\renewcommand{\r}{\mathbf{r}}
\newcommand{\h}{\mathbf{h}}
\newcommand{\w}{\mathbf{w}}
\newcommand{\g}{\mathbf{g}}
\newcommand{\I}{\mathbf{I}}

\newcommand{\X}{\mathbf{X}}
\newcommand{\s}{\mathbf{s}}
\newcommand{\f}{\mathbf{f}}
\renewcommand{\S}{\mathbf{S}}
\newcommand{\Y}{\mathbf{Y}}
\newcommand{\R}{\mathbf{R}}

\newcommand{\W}{\mathbf{W}}
\newcommand{\V}{\mathbf{V}}
\newcommand{\G}{\mathbf{G}}
\renewcommand{\H}{\mathbf{H}}
\newcommand{\z}{\mathbf{z}}

\newcommand{\Stats}{\mathbb{T}}
\newcommand{\exE}{\mathbb{E}}

\newfont{\fsc}{eusm10}                         

\DeclareMathOperator*{\argmax}{arg\,max}

\newcommand{\INPUT}{\item[\algorithmicinput]}
\newcommand{\algorithmicinput}{\textbf{Input:}}
\newcommand{\OUTPUT}{\item[\algorithmicoutput]}
\newcommand{\algorithmicoutput}{\textbf{Output:}}
\newcommand{\LINE}{ \item[]}

\begin{document}
\title{Bayesian Symbol Detection in Wireless Relay Networks via Likelihood-Free Inference}
\author{Gareth W.~Peters$^{1,2}$, Ido Nevat$^{3}$, Scott A. Sisson$^{1}$, Yanan Fan$^{1}$  and  Jinhong Yuan$^{3}$
\begin{center}
{\footnotesize {\ \textit{$^{1}$ School of Mathematics and Statistics, University of New South Wales,
Sydney, 2052, Australia; \\[0pt]
email: peterga@maths.unsw.edu.au \\[0pt]
$^{2}$ CSIRO  Sydney, Locked Bag 17,
North Ryde, New South Wales, 1670, Australia \\[0pt]
$^{3}$ School of Electrical Engineering and Telecommunications,
University of New South Wales, Sydney, Australia.} } }
\end{center}
}
%
\maketitle

\begin{abstract}
\noindent This paper presents a general stochastic model developed
for a class of cooperative wireless relay networks, in which
imperfect knowledge of the channel state information at the
destination node is assumed. The framework incorporates multiple
relay nodes operating under general known non-linear processing functions.
When a non-linear relay function is considered, the likelihood function is generally intractable resulting in the maximum likelihood and the maximum
\textit{a posteriori} detectors not admitting closed form solutions.
We illustrate our methodology to overcome this intractability under the example of a popular optimal non-linear relay function choice and demonstrate how our algorithms are capable of solving the previously intractable detection problem.
Overcoming this intractability involves development of specialised Bayesian models. We develop three novel algorithms to perform detection for this Bayesian model, these include a Markov chain Monte Carlo Approximate Bayesian Computation (MCMC-ABC) approach;
an Auxiliary Variable MCMC (MCMC-AV) approach; and a Suboptimal
Exhaustive Search Zero Forcing (SES-ZF) approach. Finally,
numerical examples comparing the symbol error rate (SER)
performance versus signal to noise ratio (SNR) of the three
detection algorithms are studied in simulated examples.
\end{abstract}



\section{Background}
Cooperative communications systems, \cite{nosratinia2004ccw} and \cite{laneman:2004}, have become a major focus for communications engineers. Particular attention has been paid to the wireless network setting, as it incorporates spatial resources to gain diversity, and enhance connection capability and throughput. More recently, the focus has shifted towards incorporation of relay nodes \cite{vandermeulen:1971}, which are known to improve energy efficiency, and reduce the interference level of wireless
channels \cite{laneman2003dst}.

In simple terms, such a system broadcasts a signal from a transmitter at the source through a wireless channel. The signal
is then received by each relay node and a relay strategy is applied before the signal is retransmitted to the destination.
A number of relay strategies have been studied in the literature (\cite{cover:1979}; \cite{laneman:2004}). We focus on the
\textit{amplify-and-forward} strategy of \cite{chen:2006} in which the relay sends a transformed version (determined by the relay function) of its received signal to the destination. The relay function can be optimized for different design objectives (\cite{kramer:2005}; \cite{khojastepour:ccr}; \cite{gomadam:2006}). It is common in the relay network literature to consider non-linear relay function choices which satisfy some concept of optimality. For example, in the estimate and forward (EF) scheme, in the case of BPSK signaling, the optimal relay function is the hyperbolic tangent \cite{ghabeli2008new}. 
Other criteria for which the optimal relay function is non-linear include: capacity maximisation \cite{ghabeli2008new}, minimum error probability at the receiver \cite{crouse2007optimal}, SNR maximisation \cite{gomadam:2006}, rate maximisation \cite{yao2008sawtooth} and minimisation of the average error probability \cite{cui2008some}. 

\begin{figure}
    \centering
        \epsfysize=6cm
        \epsfxsize=9cm
        \epsffile{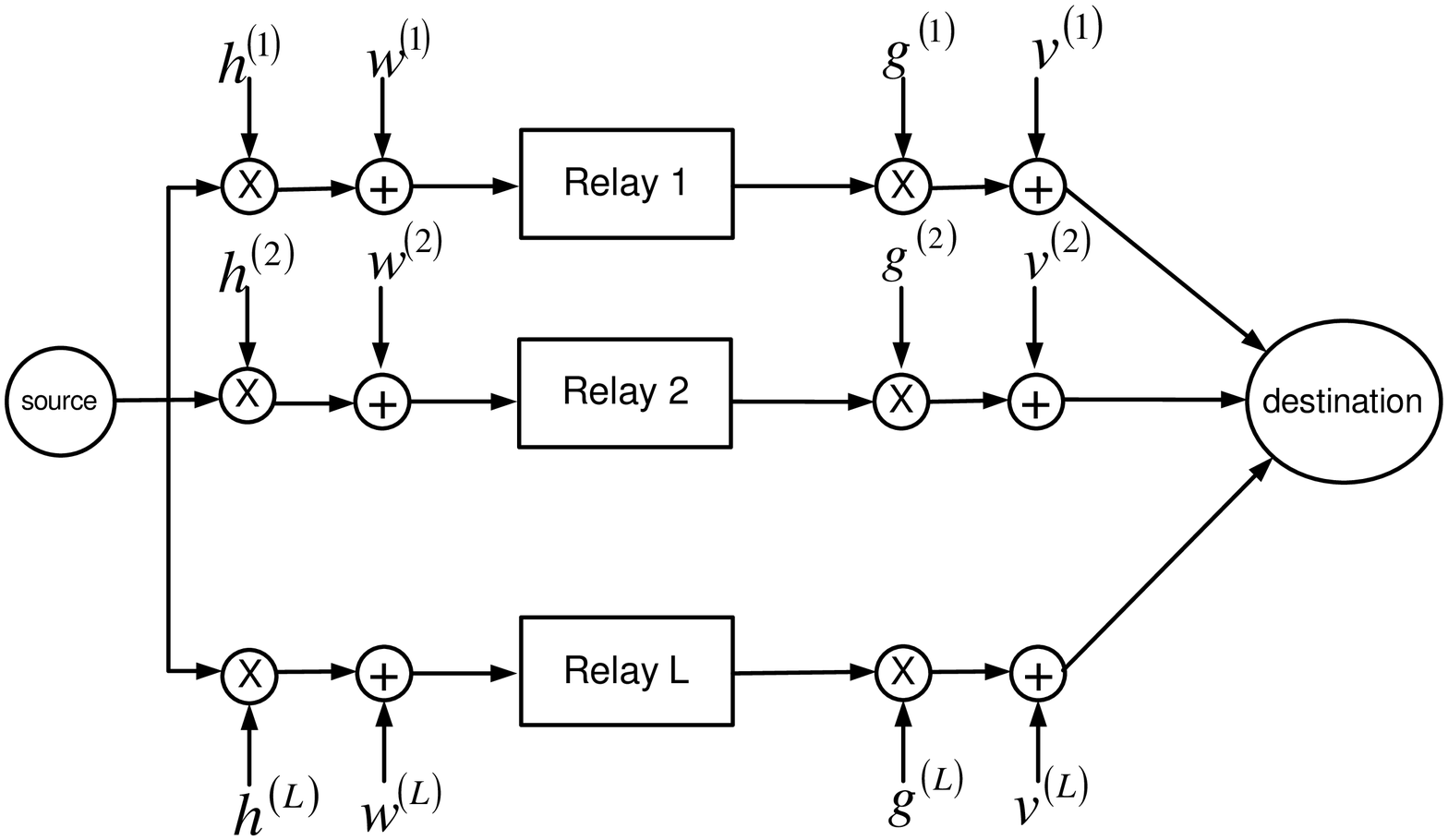}
        \caption{{\small{The system model studied in this paper involves
        transmission from one source, through each of the $L$ relay channels, $h^{(l)}$, to the relay.
        Additive complex Gaussian noise, $\W$ was included at the receiver of the relay
        then the signal is processed and retransmitted by the relay to the destination.
        In this process the signal is transmitted again through $L$ channels denoted
        by $g^{(l)}$ and additive complex Gaussian noise, $\V$ was included at the destination.}}}
    \label{fig:system}
\end{figure}

Currently, in all these cited works the authors have solved the problem of what the optimal choice of the non-linear relay function should be in order to satisfy the desired design constraint. However, we note that in all cases, whenever a non-linear relay function is considered, though it may satisfy the constraint of optimality, it will result in an intractable detection problem.
This intractability has not been considered in the literature and therefore it has not been possible to tackle the resulting detection problem. We specify explicitly how this intractability arises and then develop and demonstrate extensively our solution to this general relay network detection problem.
In this paper we will demonstrate that the choice of relay function directly affects the tractability of the system model.
We will focus on a single hop relay design in which the number of relays and the type of relay function are allowed to be known, general non-linear functions. However, our methodology trivially extends to arbitrary relay topologies and multiple hop networks. Fig. \ref{fig:system} presents the system model considered.
Our model incorporates stochastic parameters that are associated with each relay channel link. That is, we consider parameters associated with the model as random variables, which must be jointly estimated with the unknown random vector of transmitted symbols.

Since this framework incorporates general relay functions, the resulting likelihood for the model typically cannot be evaluated
pointwise, and in some cases cannot even be written down in a closed form. Bayesian "likelihood-free" inference methodologies
overcome the problems associated with the intractable likelihood by replacing explicit evaluation of the likelihood with simulation
from the model. These methods are also collectively known as Approximate Bayesian Computation (ABC), see \cite{ratmann2007ulf}, \cite{beaumont2002abc}, \cite{peters2009}, \cite{peters2008}, \cite{sisson2008}, \cite{sisson2010likelihood}, and references therein for a detailed overview of the methodological and theoretical aspects.

Applications of ABC methods are becoming widespread, and include: telecommunications  \cite{sisson2008}; extreme value theory \cite{bortot2007ise};  
protein networks, (\cite{ratmann2007ulf}, \cite{ratmann2009}); SIR models \cite{toni2009}; species migration \cite{hamilton2005}; pathogen transmission \cite{tanaka2006}, non-life insurance claims reserving \cite{peters2009}; and operational risk \cite{peters2006bim}.
\subsection{Main contributions}
In this paper we develop a novel sampling methodology based on the Markov Chain Monte Carlo Approximate Bayesian Computation
(MCMC-ABC) methodology \cite{marjoram2003mcm}. Since the focus of this
paper is on detection of the transmitted symbols, we will
integrate out the nuisance channel parameters. We also develop two novel alternative solutions for comparison: an auxiliary variable MCMC (MCMC-AV)
approach and a suboptimal explicit search zero forcing (SES-ZF)
approach. In the MCMC-AV approach the addition of auxiliary
variables results in closed form expressions for the full
conditional posterior distributions. We use this fact to develop
the MCMC-AV sampler and demonstrate that this works well when
small numbers of relay nodes are considered. The SES-ZF approach
involves an approximation based on known summary statistics of the
channel model and an explicit exhaustive search over the parameter
space of code words. This performs well for relatively small
numbers of relays and a high signal to noise ratio (SNR).

The paper is organised as follows: in Section \ref{SystemDescription} we introduce a stochastic system model for the wireless relay system and the associated Bayesian model. We discuss the intractability arising from these models when one considers arbitrary relay functions. Section \ref{ABCsection} presents the likelihood-free methodology and the generic MCMC-ABC algorithm.
In Section \ref{MCMC-ABC algorithm} we present the proposed algorithm, namely maximum \textit{a posteriori} sequence detection.
In Section \ref{AuxiliaryMCMC} we derive an alternative novel algorithm based on an auxiliary variable model for the joint
problem of coherent detection and channel estimation for arbitrary relay functions. We contrast the performance of this approach with the performance of the MCMC-ABC based algorithm. Section \ref{alternative_MAP} presents a detector that is based on known summary statistics of the channel model and an explicit exhaustive search over the parameter space of code words. We shall demonstrate
its performance, whilst acknowledging its flaws. Section \ref{simulationresults} presents results and analysis and conclusions are provided in Section \ref{conclusions}.

The following notation is used throughout: random variables are denoted by upper case letters and their realizations by lower case letters.
In addition, bold will be used to denote a vector or matrix quantity, upper subscripts will refer to the relay node index and lower subscripts refer to the element of a vector or matrix.
We define the following notation that shall be used throughout, $\g=g_{1:L}=\left(g^{(1)},\cdots, g^{(L)}\right)$; 
$\g^{[n]}$  refers to the state of the Markov chain at iteration $n$; $\g_i^{[n]}$ refers to the $i$-th element of $\g^{[n]}$.
We define the generic transition kernel for the Markov chain as $q\left(\theta^* \leftarrow \theta^{\left[n-1\right]}\right)$, which updates the Markov chain for the posterior parameters, $\theta$, from the $\left(n-1\right)$-th Markov chain step to the $n$-th via the proposal $q$.

\section{Bayesian system model and MAP detection} \label{SystemDescription}
In this section we introduce the system model and a Bayesian model for inference on the system model parameters. In our system
model, the channels in the relay network are modelled stochastically, where we do not know \textit{a priori} the
realized channel coefficient values. Instead, we consider partial channel state information (CSI), in which we assume known
statistics of the distribution of the channel coefficients.





\subsection{System model and assumptions}
Here we present the system model and associated assumptions. The system model is depicted in Fig. \ref{fig:system}.
\emph{
\begin{enumerate}
\item{Assume a wireless relay network with a single source node, transmitting sequences of $K$ symbols denoted $\s=s_{1:K}$, to a single destination via $L$ relay nodes.}
\item{The symbols in the sequence of $K$ symbols $\s$ are taken from a digital constellation with cardinality $M$.}
\item{There are $L$ relays which cannot receive and transmit on the same time slot and on the same frequency band. We thus consider a half
duplex system model in which the data transmission is divided into
two steps. In the first step, the source node broadcasts a code
word $\s \in \Omega$ from the codebook to all the $L$ relay nodes. In
the second step, the relay nodes then transmit their signals
to the destination node on orthogonal non-interfering channels. We
assume that all channels are independent with a coherence interval
larger than the codeword length $K$.}
\item{Assume imperfect CSI in which noisy estimates of the channel model
coefficients for each relay link are known. This is a standard
assumption based on the fact that a training phase has been
performed \textrm{a priori}. This involves an assumption regarding
the channel coefficients as follows:
\begin{itemize}
\item{From source to relay there are $L$ i.i.d. channels parameterized by
$\left\{H^{(l)} \sim
F\left(\widehat{h}^{(l)},\sigma_h^2\right)\right\}_{l=1}^L$, where
$F(\cdot)$ is the distribution of the channel coefficients,
$\widehat{h}^{(l)}$ is the estimated channels coefficient and
$\sigma_h^2$ is the associated estimation error variance.}
\item{From relay to destination there are $L$ i.i.d. channels parameterized by
$\left\{G^{(l)} \sim
F\left(\widehat{g}^{(l)},\sigma_g^2\right)\right\}_{l=1}^L$, where
$F(\cdot)$ is the distribution of the channel coefficients,
$\widehat{g}^{(l)}$ is the estimated channels coefficient and
$\sigma_g^2$ is the associated estimation error variance.}
\end{itemize} }
\item{ The received signal at the $l$-th relay is a random vector given by
\begin{align}
\R^{(l)} = \S H^{(l)}+ \W^{(l)}, \; l \in
\left\{1,\cdots,L\right\},
\end{align}
where $H^{(l)}$ is the channel coefficient between the transmitter
and the $l$-th relay, $\S \in \Omega_M$ is the transmitted
code-word and $\W^{(l)}$ is the noise realization associated with
the relay receiver.}
\item{The received signal at the destination is a random vector given by
\begin{align}
\label{system_model}
\Y^{(l)} = \f^{(l)}\left(\R^{(l)}\right)G^{(l)}+\V^{(l)}, \; l\in
\left\{1,\cdots,L\right\},
\end{align}
where $G^{(l)}$ is the channel coefficient between the $l$-th
relay and the receiver, \\ $\f^{(l)}\left(\r^{(l)}\right)
\triangleq
\left[f^{(l)}\left(r_1^{(l)}\right),\ldots,f^{(l)}\left(r_K^{(l)}\right)
\right]^{\top}$ is the memoryless relay processing function (with
possibly different functions at each of the relays) and $\V^{(l)}$
is the noise realization associated with the relay receiver.}
\item{Conditional on $\left\{h^{(l)},g^{(l)}\right\}_{l=1}^L$, we have that all
received signals are corrupted by zero-mean additive white complex
Gaussian noise. At the $l$-th relay the noise corresponding to the
$l$-th transmitted symbol is denoted by random variable
$W_{i}^{(l)} \sim \mbox{\fsc CN}\left(0,\sigma_w^2\right)$. At the
receiver this is denoted by random variable $V_{i}^{(l)}
\sim \mbox{\fsc CN}\left(0,\sigma_v^2\right)$. \\
Additionally, we assume the following properties:
\begin{equation*}
\mathbb{E}\left[W_{i}^{(l)} \overline{W}_{j}^{(m)} \right]=
\mathbb{E}\left[V_{i}^{(l)} \overline{V}_{j}^{(m)} \right]=
\mathbb{E}\left[W_{i}^{(l)} \overline{V}_{j}^{(m)} \right]=0,
\end{equation*}
$\forall i,j \in \left\{1,\ldots,K\right\}, \forall l,m \in
\left\{1,\ldots,L\right\}, i \neq j, l \neq m$, where
$\overline{W}_{j}$ denotes the complex conjugate of $W_j$.}
\end{enumerate}
}

\subsection{Prior specification and posterior}
Here we present the relevant aspects of the Bayesian model and associated assumptions. 
In order to construct a Bayesian model we need to specify the likelihood $p\left(\y|\s,\h,\g\right)$, and the priors for the parameters $\s,\h,\g$ which are combined under Bayes theorem to obtain a posterior $p\left(\s,\g,\h|\y\right)$.
We begin by specifying the prior models
for the sequence of symbols and the unknown channel coefficients.
At this stage we note that in terms of system capacity it is only
beneficial to transmit a sequence of symbols if it aids detection.
This is achieved by having correlation in the transmitted symbol sequence $s_{1:K}$. We assume that since this is part of the system
design, the prior structure for $p\left(s_{1:K}\right)$ will be
known and reflect this information. \emph{
\begin{enumerate}
\item{Under the Bayesian model, the symbol sequence is treated
as a random vector $\S = S_{1:K}$. The prior for the random
symbols sequence (code word) $S_{1:K}$ is defined on a discrete
support denoted $\Omega$ with $\left|\Omega\right|=M^K$
elements and probability mass function denoted by
$p\left(s_{1:K}\right)$.}
\item{The assumption of imperfect CSI is treated under a Bayesian paradigm
by formulating priors for the channel coefficients as follows:
\begin{itemize}
\item{From source to relay there are $L$ i.i.d. channels parameterized by
$\left\{H^{(l)} \sim \mbox{\fsc
CN}\left(\widehat{h}^{(l)},\sigma_h^2\right)\right\}_{l=1}^L$,
where $\widehat{h}^{(l)}$ is the estimated channels coefficient
and $\sigma_h^2$ the associated estimation error variance.}
\item{From relay to destination there are $L$ i.i.d. channels parameterized by
$\left\{G^{(l)} \sim \mbox{\fsc
CN}\left(\widehat{g}^{(l)},\sigma_g^2\right)\right\}_{l=1}^L$,
where $\widehat{g}^{(l)}$ is the restituted channels coefficient
and $\sigma_g^2$ the associated estimation error variance.}
\end{itemize} }
\end{enumerate}
}
\subsection{Inference and MAP sequence detection} \label{non linear MAP detection}
Since the primary concern in designing a relay network system is
on SER versus SNR, our goal is oriented towards detection of
transmitted symbols and not the associated problem of channel
estimation. We will focus on an approach which samples
$S_{1:K},\G=G_{1:L},\H=H_{1:L}$ jointly from the target posterior
distribution, since our model also considers the channels to be stochastic and unknown.

In particular we consider the maximum \textit{a posteriori} (MAP)
sequence detector at the destination node. Therefore the goal is
to design a MAP detection scheme for $\s$ at the destination,
based on the received signals $\left\{\y^{(l)}\right\}^L_{l=1}$,
the noisy channels estimates as given by the partial CSI
$\left\{\widehat{h}^{(l)}\right\}^L_{l=1}$, $\left\{\widehat{g}^{(l)}\right\}^L_{l=1}$, $\sigma_h^2$, $\sigma_g^2$, and the noise variances, $\sigma^2_w$ and $\sigma^2_v$.

Since the channels are mutually independent, the received signals
$\left\{\r^{(l)}\right\}^L_{l=1}$ and
$\left\{\y^{(l)}\right\}^L_{l=1}$ are conditionally independent
given $\s,\g,\h$. Thus, the MAP decision rule, after marginalizing
out the unknown channel coefficients, is given by
\small
\begin{align}
\begin{split}
\label{MAP_detector} 
\widehat{\s} &= \argmax_{\s \in \Omega} p\left(\s|\y\right) \\
&=\argmax_{\s \in \Omega}
\prod^L_{l=1}\int \int p\left(\s,g^{(l)},h^{(l)}|\y^{(l)}\right) d\h d\g\\
&=\argmax_{\s \in \Omega} \prod^L_{l=1} \int \int
p\left(\y^{(l)}|\s,h^{(l)},g^{(l)}\right)p\left(\s\right)
p\left(\g\right)
p\left(\h\right)d\h d\g.
\end{split}
\end{align}
\normalsize
Therefore, in order to perform detection of the transmitted symbols, we need to evaluate the likelihood function of the model.
In the next section we will demonstrate that the likelihood function in our model is intractable and as a result we develop the likelihood-free based methodology and associated MCMC-ABC algorithm to perform the MAP detection. 
\subsection{Evaluation of the likelihood function}
The likelihood model $p\left(\y^{(l)}|\s,h^{(l)},g^{(l)}\right)$
for this relay system is in general computationally intractable as we now show.
There are two potential difficulties that arise when dealing with
non-linear relay functions. The first relates to finding the
distribution of the signal transmitted from each relay to the
destination. This involves finding the density of the random
vector $\f^{(l)}\left(\S H^{(l)} +\W^{(l)}\right)G^{(l)}$
conditional on realizations $\S =\s,\H=\h,\G=\g$. This is not
always possible for a general non-linear multivariate function
$\f^{(l)}$. Conditional on $\S =\s,\H=\h,\G=\g$, we know the
distribution of $\R^{(l)}|\s,\g,\h$,
\begin{align*}
\begin{split}
& p_{\R}\left(\r|\s,\g,\h\right) \\
&=p \left(\s
h^{(l)}+\w^{(l)}|\s,h^{(l)},g^{(l)}\right) =\mbox{\fsc CN}\left(
\s h^{(l)},\sigma^2_w \I \right).
\end{split}
\end{align*}
However, finding the distribution of the random vector after the
non-linear function is applied i.e. the distribution of $
\widetilde{\f}\left(\R^{(l)}\right)  \triangleq
\f\left(\R^{(l)}\right) G^{(l)}$ given $\s,h^{(l)},g^{(l)}$,
involves the following change of variable formula
\begin{align*}
\begin{split}
&p\left(\widetilde{\f}\left(\r^{(l)}\right)|\s,h^{(l)},g^{(l)}\right)\\
&=
p_{\R}\left((\widetilde{\f}^{(l)})^{-1}\left(\r^{(l)}\right)|\s,h^{(l)},g^{(l)}
\right) \left|\frac{\partial \widetilde{\f}^{(l)}}{\partial
\r^{(l)}}\right|^{-1},
\end{split}
\end{align*}
which can not always be written down analytically for arbitrary
$\widetilde{\f}$. The second more serious complication is that
even in cases where the density for the transmitted signal is
known, one must then solve a $K$-fold convolution to obtain the
likelihood:
\begin{align}
\label{likelihood}
\begin{split}
&p\left(\y^{(l)}|\s,\g,\h\right)=p\left(\widetilde{\f}\left(\r^{(l)}\right)|\s,\g,\h\right) \ast p_{\V^{(l)}} \\
&= \int^{\infty}_{-\infty}\cdots \int^{\infty}_{-\infty}
p\left(\widetilde{\f}\left(\z|\s,\g,\h\right)\right)
p_{\V^{(l)}}\left(\y^{(l)}-\z\right) d z_1 \ldots d z_K,
\end{split}
\end{align}

where $\ast$ denotes the convolution operation.
Typically this will be intractable to evaluate pointwise. However,
in the most simplistic case of a linear relay function, that is, $\f^{(l)}\left(x \right)=x$, the likelihood can be obtained analytically as
\begin{align*}
\label{optimal_map} &p\left(\y^{(l)}|\s,h^{(l)},g^{(l)}\right)=
\mbox{\fsc CN}\left(\s h^{(l)} g^{(l)}
,\left(\left|g^{(l)}\right|^2 \sigma_w^2
+\sigma_v^2\right)\I\right),
\end{align*}
where $\I$ is the identity matrix.

Hence, the resulting posterior distribution in (\ref{MAP_detector}) involves combining the
likelihood given in (\ref{likelihood}) with the priors for
the sequence of symbols and the channel coefficients.
\section{Likelihood-free methodology} \label{ABCsection}
In this section, we provide a concise background introduction on likelihood-free methodology, also known as approximate Bayesian computation, and the sampling algorithms utilised to obtain samples from the ABC posterior model, based on \cite{sisson2008}, \cite{toni2009}, \cite{marjoram2003mcm}, \cite{peters2008}.
Likelihood-free inference describes a suite of methods developed
specifically for working with models in which the likelihood is
computationally intractable. 
We consider the likelihood intractability to arise in the sense that we may not evaluate the likelihood pointwise, such as in (\ref{likelihood}). In particular, for the relay models considered, we can only obtain a general expression for the likelihood in terms of multivariate convolution integrals and hence we do not have an explicit closed form expression for which to evaluate the likelihood pointwise.\\
\textbf{ABC Methodology and MCMC-ABC:}		\\
All ABC methodologies are based upon the observation that one may replace the problem of evaluating the intractable likelihood point-wise with the typically trivial exercise of simulating from the likelihood model. Simulation from the likelihood usually involves trivial generation of random variables from a known parametric distribution, followed by application of a transformation corresponding to the imposed physical model. 

It is shown in \cite{reeves2005} and further developed in \cite{sisson2008} that the ABC method embeds an ``intractable" target posterior distribution (resulting from the intractability of the likelihood function (\ref{likelihood})), denoted by $p\left(\s|\y\right)$,
into an augmented posterior model,
\begin{equation*}
p\left(\s,\x|\y\right) \propto
p\left(\y|\x,\s;\epsilon\right)p\left(\x|\s\right)p\left(\s\right),
\end{equation*}
where $\x \in \mathbb{X}$ is an auxiliary vector on the same
space as observations $\y$. In this augmented Bayesian model, the density $p\left(\y|\x,\s;\epsilon \right)$ acts as a weight for the
intractable posterior. It compares the observations $\y$ with the auxiliary ("synthetic data" $\x$) generated from the likelihood model, and determines whether or not they are within an $\epsilon$ tolerance of each other. Choices for this weighting, $p\left(\y|\x,\s;\epsilon \right)$, will be explained below. For detailed justification for this framework see \cite{sisson2008}, \cite{peters2009} and \cite{marjoram2003mcm}.  
In this paper we consider marginal sampler. Summarising the ideas in these papers, we note that the target marginal posterior $p(\s|\y)$ can be represented in the ABC framework as follows
\begin{align}
\label{ABC_framework}
\begin{split}
p\left(\s|\y\right) &\propto p\left(\s\right) p\left(\y|\s\right)\\
&=p\left(\s\right) \int_{\mathbb{X}} p(\y|\x,\s; \epsilon)p(\x|\s)d\x\\
&=  p\left(\s\right) \exE_{p(\x|\s)} \left[p(\y|\x,\s; \epsilon)\right]\\
&\approx \frac{p\left(\s\right)}{D}  \sum_{d=1}^D p(\y|\x^{d},\s; \epsilon) := \widehat{p}\left(\s|\y\right).
\end{split}
\end{align}
Presenting the ABC approximate posterior in this way allows us to provide intuition for the ABC mechanism. To understand this, we note that the Monte Carlo approximation, denoted by $\widehat{p}\left(\s|\y\right)$, of the integral with respect to the intractable likelihood, is approximated
via $D$ draws of auxiliary variables $\x^d$ ("synthetic data") from $p(\x|\s)$. Therefore, this demonstrates explicitly how one can replace the evaluation of the likelihood with simulation from the likelihood model. As discussed in \cite{peters2009}, \cite{peters2008}, the choice of weighting function can affect the Monte Carlo estimate.
The function $p(\y|\x,\s; \epsilon)$ is typically a standard smoothing kernel \cite{blum2009approximate} with scale parameter $\epsilon$ which weights the intractable posterior with high values in regions when the observed data $\y$ and auxiliary date $\x$ are similar. For example, uniform kernels are commonplace in likelihood-free models (e.g. \cite{marjoram2003mcm}), although alternatives such as Epanechnikov \cite{beaumont2002abc} and Gaussian kernels \cite{peters2008} provide improved efficiency.

In this paper we consider two popular choices for weighting functions, $p(\y|\x,\s; \epsilon)$, which are easily interpreted as hard and soft decision functions \cite{peters2008}.
The popular "Hard Decision" (HD) is given by 
\begin{align}
\label{WeightingFunctionHD}
p\left(\y|\x,\s;\epsilon\right) &\propto
\begin{cases}
{1,} & \text{if }\rho\left(\y,\x\right)\leq\epsilon, \\
{0,} & \text{ otherwise.}%
\end{cases}
\end{align}
This makes a hard decision to reward those simulated auxiliary variables, $\x$, that are within an
$\epsilon$-tolerance of the actual observed data, $\y$, as measured by distance metric $\rho$. 
Another example is a "Soft Decision" (SD) weighting function given by 
\begin{align}
\label{WeightingFunctionSD}
p\left(\y|\x, \s ;\epsilon\right)
&\propto
\exp\left(-\frac{\rho\left(\y,\x\right)}{\epsilon^2}\right).
\end{align}

\noindent An important practical difference between the HD and SD kernels is that, even though the weighting of the intractable posterior, obtained by the SD density, may be small, it will remain non-zero unlike the HD rule. 

The choice of $\epsilon$ is therefore important for performances of the ABC methodology. If $\epsilon$ is too large, the approximation $\widehat{p}\left(\s|\y\right)$ is poor; for example when $\epsilon \rightarrow \infty$ the resulting ABC posterior $\widehat{p}\left(\s|\y\right)$ in (\ref{ABC_framework}) corresponds to the prior since the weighting function is uniform for all values of the posterior parameters. However, if $\epsilon$ is sufficiently small, $\widehat{p}\left(\s|\y\right)$ is a good approximation of $p\left(\s |\y\right)$. We note that there is no approximation when $\epsilon=0$. From a practical computational cost perspective, we will discuss why it is not feasible to set $\epsilon = 0$ and instead we must settle for selecting an $\epsilon > 0$ via a trade-off between accuracy and computation cost. There are several guidelines to specify a sequence of tolerances that one may follow to achieve a given $\epsilon$, see \cite{del2009adaptive}, \cite{peters2008}, \cite{ratmann2009}.

The following two sections provide specific background detail related to specifications of the weighting functions. We then finish this section by presenting the generic MCMC-ABC algorithm used to obtain samples from the marginal posterior $p\left(\s|\y\right)$ in order to solve the MAP detection problem in (\ref{MAP_detector}).
\subsection{Data Summaries} \label{Data_Summaries}
We note that when the total number of observations is large, then comparing directly the simulated data $\x$ to the observed data $\y$ in the weighting function is not computationally efficient. Therefore, one considers dimension reduction. The standard approach to dimension reduction is to compare summary statistics, $\Stats\left(\y\right)$, which should summarise the information present in the data. 
The summary statistic $\Stats\left(\y\right)$ is called a sufficient statistic for $\y$ if and only if
$ p\left(\s|\y\right)\propto p\left(\s|\Stats\left(\y\right)\right),$ i.e., according to the Neayman-Fisher factorisation theorem, we can replace the conditional distribution of $\y$ given the parameters with the conditional distribution summary statistics given the parameters and a constant with respect to the parameters.
If a sufficient statistic is available, then using the summary statistics is essentially the same as using the whole data.
Therefore, as $\epsilon \rightarrow 0$, the ABC approximation $\widehat{p}(\s|\Stats\left(\y\right))\rightarrow p(\s|\y))$.
 However, in most real practical models, sufficient statistics are unknown and one must make alternative choices.
\subsection{Distance Metrics}

Having obtained summary statistic vectors $\Stats\left(\y\right)$ and $\Stats\left(\x\right)$, likelihood-free methodology then
measures the distance between these vectors using a distance metric, denoted generically by  $\rho\left(\Stats\left(\y\right),\Stats\left(\x\right)\right)$. The most popular example involves the basic Euclidean distance metric which sums up the squared error between each summary statistic as follows:
\begin{equation}
\rho\left(\Stats\left(\y\right),\Stats\left(\x\right)\right)=
\sum _i^{\text{dim}\left(T\right)} \left(\Stats_i\left(\y\right)-\Stats_i\left(\x\right)\right)^2.
\end{equation}
Recently more advanced choices have been proposed and their impact on the methodology has been assessed \cite{peters2008}. These include: scaled Euclidean distance given by the square root of
\begin{align}
\rho\left(\Stats\left(\y\right),\Stats\left(\x\right)\right)=
\sum _i^{\text{dim}\left(T\right)} \mathbf{\Lambda}_i \left(\Stats_i\left(\y\right)-\Stats_i\left(\x\right)\right)^2;
\end{align}
Mahalanobis distance:
\begin{eqnarray}
\rho\left(\Stats\left(\y\right),\Stats\left(\x\right)\right)=
\left(\Stats\left(\y\right)-\Stats\left(\x\right)\right) \Sigma^{-1}
\left(\Stats\left(\y\right)-\Stats\left(\x\right)\right)^T; 
\end{eqnarray}
$L^p$ norm:
\begin{eqnarray}
\rho\left(\Stats\left(\y\right),\Stats\left(\x\right)\right)=
\sum _i^{\text{dim}\left(T\right)}
\left[\left|\Stats_i\left(\y\right)-\Stats_i\left(\x\right)\right|^p\right]^{1/p};
\end{eqnarray}
and city block distance:
\begin{eqnarray}
\rho\left(\Stats\left(\y\right),\Stats\left(\x\right)\right)=
\sum _i^{\text{dim}\left(T\right)}\left|\Stats_i\left(\y\right)-\Stats_i\left(\x\right)\right|.
\end{eqnarray}
In particular, we note that distance metrics which include information regarding correlation
between elements of each summary statistic vector, result in improved estimates of the marginal posterior.
\subsection {MCMC-ABC Samplers}		

MCMC-based likelihood-free samplers were introduced to avoid the basic rejection sampling algorithms (see for example \cite{plagnol2004approximate}) which are inefficient when the posterior and prior are sufficiently different \cite{marjoram2003mcm}. Therefore, an MCMC approach that is based on the ABC methodology has been devised \cite{marjoram2003mcm}.
The generic MCMC-ABC sampler is presented in Algorithm \ref{abc_mcmc_sampler}.
\begin{algorithm}
 \caption{generic MCMC-ABC sampler}
 \label{abc_mcmc_sampler}
	 \begin{algorithmic}[1]
    \INPUT $p\left(\s\right)$
    \OUTPUT $\left\{\s^{\left[n\right]}\right\}_{n=1}^N$, samples from $p\left(\s|\y \right) $
     \STATE Initialise $\s^{\left[1\right]}$ to an arbitrary staring point
     \FOR{$n=2,\hdots, N$}
    \STATE Propose $\s^{*} \sim q(\s^{*} \leftarrow \s^{\left[n-1\right]})$
    \STATE Simulate synthetic data from the model, $\X \sim p\left(\x|\s^*\right)$
    \IF{ $\rho\left(\Stats\left(\y\right),\Stats\left(\x\right)\right) \leq \epsilon$} 
        \STATE Compute the MCMC acceptance probability:\\
         $\alpha\left(\s^{*},\s^{\left[n-1\right]}\right)
        =\min\left\{1,\frac{p\left(\s^{*}\right)}{p\left(\s^{\left[n-1\right]}\right)} \times
           		\frac{q(\s^{\left[n-1\right]} \leftarrow \s^{*}) }{q(\s^{*} \leftarrow \s^{\left[n-1\right]})}\right\}$
    			\STATE Generate $u \sim U\left[ 0,1 \right]$  			
    			\IF {$u \leq \alpha\left(\s^{*},\s^{\left[n-1\right]}\right) $}
    			    \STATE $\s^{\left[n\right]}=\s^{*}$
    			        \ELSE
    			     \STATE $\s^{\left[n\right]}=\s^{\left[n-1\right]}$ 
    			\ENDIF    	
    \ELSE
    			\STATE $\s^{\left[n\right]}=\s^{\left[n-1\right]}$ 
    \ENDIF
    \ENDFOR
  \end{algorithmic}
\end{algorithm}

In the next section we present details of choices that must be made when constructing a
likelihood-free inference model.

\section{MCMC-ABC Based Detector }\label{MCMC-ABC algorithm}
We now relate the MCMC-ABC generic sampler to the specific problem presented in this paper. As mentioned previously, the ABC method we consider embeds an intractable target posterior distribution, in our
case denoted by $p\left(s_{1:K},h_{1:L},g_{1:L}|\y\right)$, into a
general augmented model
\begin{align}
\begin{split}
&p\left(s_{1:K},h_{1:L},g_{1:L},\x|\y\right) \propto
p\left(\y|\x,s_{1:K},h_{1:L},g_{1:L}\right) \\
&p\left(\x|s_{1:K},h_{1:L},g_{1:L}\right)
p\left(s_{1:K}\right)p\left(h_{1:L}\right)p\left(g_{1:L}\right),
\end{split}
\end{align}
where $\x$ is an auxiliary vector on the same space as $\y$. In this paper we make the standard ABC assumption, for the weighting function,
$p\left(\y|\x,s_{1:K},h_{1:L},g_{1:L}\right) = p\left(\y|\x\right)$, see \cite{ratmann2009}.

Hence, in the ABC context, we obtain a general approximation to the intractable full posterior, denoted by
$\widehat{p}\left(s_{1:K},h_{1:L},g_{1:L}|\y\right)$. We are interested in the marginal target posterior, $p\left(s_{1:K}|\y\right)$, which, in the ABC framework is approximated by
\small
\begin{align}
\begin{split}
\label{postABC}
& p\left(s_{1:K}|\y \right) \\
&\propto \int \int \int
p\left(\y|\x,s_{1:K},h_{1:L},g_{1:L};\epsilon\right)p\left(\x|s_{1:K},h_{1:L},g_{1:L}
\right)p\left(s_{1:K}\right)\\
&\;\;\;\;\;\;\;\;\;\;\;\;\;\;\;\;\;p\left(h_{1:L}\right)p\left(g_{1:L}\right)d\h
d\g d\mathbf{x}\\
&\approx 
p\left(s_{1:K}\right)
\sum_{n=1}^N \sum_{d=1}^D
p\left(\y|\x^{d,\left[n\right]},s_{1:K},h^{\left[n\right]}_{1:L},g^{\left[n\right]}_{1:L};\epsilon\right)p\left(h^{\left[n\right]}_{1:L}\right)p\left(g^{\left[n\right]}_{1:L}\right)\\
&:=\widehat{p}\left(s_{1:K}|\y \right),
\end{split}
\end{align}
\normalsize
where $\x^{d,\left[n\right]}$ represents the $d$-th realisation at the $n$-th step of the Markov chain.
In this paper we are interested in the marginal posterior ABC approximation $\widehat{p}\left(s_{1:K}|\y\right)$, as we wish to formulate the MAP detector for the symbols.

Next we present the resulting MCMC-ABC algorithm to perform MAP detection of a
sequence of transmitted symbols. 
In particular our MCMC sampler is comprised of a random scan Metropolis-Hastings (MH) within
Gibbs sampler \cite{chib1995understanding}. 
The Gibbs sampler is a special case of the MH algorithm. It is typically used when one wishes to update sub-blocks of the posterior parameters at each iteration of the Markov chain. This is especially beneficial when the full conditional posterior distributions for each sub-block can be expressed in closed form expressions and sampled. In this case the transition kernel for the Markov chain on the full set of posterior parameters becomes a product of the full conditional posterior densities, resulting in acceptance probability of one, see \cite{gilks1996markov}.
If however, one cannot sample easily from any of the conditional posterior densities, an MH algorithm is used to produce a sample. This algorithm is referred to as a Metropolis-Hastings within Gibbs sampler.

Algorithm \ref{MCMC-ABC algo} presents the details of the sampler, where we use the compact notation $\Theta =
\left(\S,\G,\H\right)$.
\begin{algorithm}
 \caption{MAP sequence detection algorithm using MCMC-ABC}
 \begin{algorithmic}[1]
\label{MCMC-ABC algo}
\LINE{\textbf{Initialize Markov chain state:}} 
\STATE Initialize
n=1, $\S^{\left[1\right]} \sim p\left(\s\right)$,
$g^{[1]}_{1:L} = \widehat{g}_{1:L}$,
$h^{[1]}_{1:L} = \widehat{h}_{1:L}$
\FOR{$n=2,\hdots,N$}
\LINE{\textbf{Propose new Markov chain state: $\Theta^{*}$ given
$\Theta^{[n-1]}$.}} \STATE Draw an index $i \sim
U\left[1,\ldots,K+2L\right]$ 
\STATE Draw proposal\\
$\bm{\Theta}^*=\left[\theta^{[n-1]}_{1:i-1},\theta^*,\theta^{[n-1]}_{i+1:K+2L}\right]$
from proposal distribution
$q(\theta^* \leftarrow \theta_i^{[n-1]})$.
(Note the proposal will depend on which element of the $\Theta$ vector is being sampled.)
\LINE{\textbf{ABC posterior (\ref{postABC}):}}
\STATE Generate auxiliary variables
$x^{\left(l\right)}_1,\ldots,x^{\left(l\right)}_K$ from the model,
$p\left(\x^{\left(l\right)}|\bm{\theta}^*\right)$, for
$l=1,\ldots,L$, to obtain a realization of  
 $\X =\x =
\left[\x^{\left(1\right)},\ldots,\x^{\left(L\right)}\right]^{\top} $ by:
\LINE (5.a)\small  \hspace{0.2cm} Sample $\W^{\left(l\right)*} \sim  \mbox{\fsc CN}\left(\textbf{0},\sigma_w^2 \I\right), \hspace{0.1cm} l\in \left\{1,\cdots,L\right\}$. 
\LINE (5.b)\small  \hspace{0.25cm} Sample $\V^{\left(l\right)*} \sim \mbox{\fsc CN}\left(\textbf{0},\sigma_v^2 \I\right), \hspace{0.1cm} l\in \left\{1,\cdots,L\right\}$.
\LINE (5.c)\small   \hspace{0.2cm} Evaluate $\X^{\left(l\right)} =
\f^{(l)}\left(\S^*h^{(l)*}+\W^{\left(l\right)*}\right)g^{(l)*}$ 
\LINE \small   \hspace{2.7cm} $+\V^{(l)*},
l\in \left\{1,\cdots,L\right\}$.
\normalsize
\STATE Calculate a measure of distance $\rho\left(\Stats(\y),\Stats(\x)\right)$

\STATE Evaluate the acceptance probability\\
\small
                \begin{equation*}
                \begin{split}
                &\alpha\left(\Theta^{[n-1]},\Theta^*\right) = \min\left\{1,
                \frac{\widehat{p}\left(\bm{\theta}^*|\y,\epsilon_n \right)q(\bm{\theta}^* \rightarrow \bm{\theta}^{[n-1]})}
                {\widehat{p}\left(\bm{\theta}^{[n-1]}| \y,\epsilon_{n-1} \right) q(\bm{\theta}^{[n-1]} \rightarrow \bm{\theta}^*)}\right\},
                \end{split}
                \end{equation*}
\normalsize                
where $p_{ABC}\left(\bm{\theta}^*|\y,\epsilon_n \right)$, depending whether HD or SD is used, is given by:\\
\small
\begin{align*}
\begin{split}
\text{HD\,: \;\;}
&\widehat{p}\left(\bm{\theta}^*|\y,\epsilon_n \right)  
\\
&\propto
\begin{cases}
p\left(s^*_{1:K}\right)p\left(h^*_{1:L}\right)p\left(g^*_{1:L}\right),
 & \text{if }	\rho\left(\Stats(\y),\Stats(\x)\right)		\leq\epsilon_n, \\
0, & \text{ otherwise;}
\end{cases}
\end{split}
\end{align*}

\begin{align*}
\begin{split}
\text{SD\,: \;\;}
&\widehat{p}\left(\bm{\theta}^*|\y,\epsilon_n \right)  \\
&\propto
\exp\left(-\frac{\rho\left(\Stats(\y),\Stats(\x)\right)}{\epsilon_n^2}\right)
p\left(s^*_{1:K}\right)p\left(h^*_{1:L}\right)p\left(g^*_{1:L}\right).
\end{split}
\end{align*}
\normalsize
\STATE Sample random variate $u$, where $U \sim
U\left[0,1\right]$.

\IF { $u \leq \alpha\left(\bm{\Theta}^{[n-1]},\bm{\Theta}^*\right)$}
\STATE $\bm{\Theta}^{[n]}=\bm{\Theta}^*$ 
\ELSE 
\STATE  $\bm{\Theta}^{[n]}=\bm{\Theta}^{[n-1]}$.
\ENDIF
\ENDFOR

\end{algorithmic}
\end{algorithm}
In order to utilise the Metropolis-Hastings algorithm, we now specify the Markov chain transition kernels for $\S,\G,\H$, otherwise known as the proposal distributions for the sampler.
We design the sampler to update a single component of the posterior parameters, at every iteration of the algorithm, as it allows us to design proposals that insure a reasonable acceptance probability in the Markov chain rejection step.
The transition kernels, denoted generically by $q\left(\theta^* \leftarrow \theta^{\left[n-1\right]}\right)$, which updates the Markov chain for the posterior parameters, $\theta$, from the $\left(n-1\right)$-th Markov chain step to the $n$-th via the proposal $q$. In this case we decompose $q$ into two components: the first, denoted by $q(i)$, specifies the transition probability mass function for sampling a proposed element of the posterior parameter vector to be updated; and the second, denoted by $q\left(\theta_i|\theta_i^{\left[n-1\right]}\right)$, specifies the probability density for proposing a new Markov chain state for element $\theta_i$ conditional on its previous state at iteration $n-1$. 
The specific transition kernels we specified for our algorithm are given by:
\begin{itemize}
\item{Transition kernel for $\S_{1:K}$: draw a proposal $\S_{1:K}^*$ from distribution
\begin{align}
\begin{split}
\label{MH_prop1}
q(\s_{1:K}^* \leftarrow \s_{1:K}^{ [n-1]})&= q(i) q(\s_i|\s_i^{[n-1]})\\
&=\frac{1}{K}\frac{1}{\log_2M}\delta_{\s_{1:i-1}^{[n-1 ]}}\delta_{\s_{i+1:K}^{[n-1]}}.
\end{split}
\end{align}
}
\normalsize
\item{Transition kernel for $\G_i$: draw proposal $\G_i^*$ from distribution 
\small
\begin{align}
q( \g^* \leftarrow \g_i^{[n-1]}) = q(l) q(g_i^{(l)}|\g_i^{[n-1]}) = \frac{1}{L} \mbox{\fsc CN} \left(\g_i^{[n-1]}, \sigma^2_{g\_rw}\right).
\end{align}
}
\normalsize
\item{Transition kernel for $\H_i$: draw proposal $\H_i^*$ from distribution
\small
\begin{align}
\label{MH_prop2}
 q( \h^* \leftarrow \h_i^{[n-1]} )
= q(l) q(h_i^{(l)}|\h_i^{[n-1]}) = \frac{1}{L}
\mbox{\fsc CN} \left(\h_i^{[n-1]},\sigma^2_{h\_rw}\right),
\end{align}
}
\normalsize
\end{itemize}
where $\delta_{\phi}$ denotes a dirac mass on location $\phi$, and
$q(i)$ and $q(l)$ respectively denote the uniform probabilities of
choosing indices $i \in \{1,\ldots,K\}$ and $l \in \{1,\ldots,L\}$, and $\sigma^2_{g\_rw}, \sigma^2_{h\_rw}$ represent the transition kernels variance.
Thus, in every iteration we update a single component of the symbols $\s_{1:K}^{ [n]}$, a single component of the transmitter-relay channels $\g_i^{[n]}$, and a single component of the relay-receiver channels $\h_i^{[n]}$.


We now present details about the choices made for the ABC components of
the likelihood-free methodology.

\subsection{Observations and synthetic data}

The data $\y=y_{1:K}$ correspond to the observed sequence of
symbols at the receiver. The generation of the synthetic data in
the likelihood-free approach involves generating auxiliary
variables $x^{\left(l\right)}_1,\ldots,x^{\left(l\right)}_K$ from
the model, $p\left(\x^{\left(l\right)}|\s,\g,\h\right)$, for
$l=1,\ldots,L$, to obtain a realization $\x =
\left[\x^{\left(1\right)},\ldots,\x^{\left(L\right)}\right]^{\top}$.
This is achieved under the following steps:
\begin{enumerate}
\item [a.] Sample $\W^{\left(l\right)*} \sim  \mbox{\fsc CN}\left(\textbf{0},\sigma_w^2 \I\right)$.
\item [b.] Sample $\V^{\left(l\right)*} \sim \mbox{\fsc CN}\left(\textbf{0},\sigma_v^2 \I\right)$
\item [c.] Evaluate the system model in (\ref{system_model}),
\small
\begin{align*}
 \X^{\left(l\right)} = \f^{(l)}\left(\S^*h^{(l)*}+\W^{\left(l\right)*}\right)g^{(l)*}+\V^{(l)*}, l\in \left\{1,\cdots,L\right\}.
 \end{align*}
 \normalsize
\end{enumerate}
\subsection{Summary statistics}
As discussed in Section \ref{Data_Summaries}, summary statistics
$\Stats(\cdot)$ are used in the comparison between the synthetic data
and the observed data via the weighting function. 
There are many possible choices of summary statistic. Most
critically, we want the summary statistics to be as close to sufficient statistics as possible
whilst low in dimension. The simplest choice is to use
$\Stats\left(\y\right)=\y$ i.e. the complete dataset. This is optimal in
the sense that it does not result in a loss of information from summarising the observations. The reason this choice is rarely used is that
typically it will result in poor performance of the MCMC-ABC as discussed previously.
 To understand this, consider the HD rule
weighting function in  (\ref{WeightingFunctionHD}). In this
case, even if the true MAP estimated model parameters were utilized
to generate the synthetic data $\x$, it will still become
improbable to realise a non-zero weight as $\epsilon \rightarrow
0$. This is made worse as the number of observations increases,
through the curse of dimensionality. As a result, the acceptance
probability in the MCMC-ABC algorithm would be zero for long periods,
resulting in poor sampler efficiency. See \cite{peters2006bim} for a
related discussion on chain mixing. We note that the exception to
this rule is when a moderate tolerance $\epsilon$ and small number
of observations are used.

A popular and practical alternative to utilizing the entire data set,
is to use empirical quantile estimates of the data distribution.
Here we adopt the vector of quantiles $\Stats\left(\y\right)=
\left[\widehat{q}_{0.1}\left(\y\right),\ldots,\widehat{q}_{0.9}\left(\y\right)
\right]$, where $\widehat{q}_{\alpha}\left(\y\right)$ denotes the
$\alpha$-level empirical quantile of $\y$, see for example \cite{bortot2007ise}, \cite{tanaka2006}.

\subsection{Distance metric}
For the distance metric $\rho\left(\Stats(\y),\Stats(\x)\right)$, as a
component of the weighting function, we use the Mahalanobis
distance
\[
\rho \left( \Stats\left(\y\right),\Stats\left(\x\right)
\right)= \left[ \Stats\left(\y\right) -\Stats \left(\x\right) \right] ^{\top
} ~\Sigma _{\x}^{-1}~ \left[ \Stats\left(\y\right) -\Stats \left(\x\right)
\right],
\]
where $\Sigma _{\x}$ is the covariance matrix of $\Stats(\x)$. 
In Section \ref{simulationresults} we contrast this distance metric with
scaled Euclidean distance, obtained by substituting
$\text{diag}\left(\Sigma_{\x}\right)$ for $\Sigma_{\x}$ in the above.

We estimate $\Sigma_{\x}$ as the sample covariance matrix of
$\Stats(\x)$, based on $2,000$ likelihood draws from
$p\left(\y|\tilde{s}_{1:K},\widehat{h}_{1:L},\widehat{g}_{1:L}\right)$
where $\tilde{s}_{1:K}$ is a mode of the prior for the symbols and
the channel coefficients are replaced with the partial CSI
estimates.

We note that in principal, the choice of matrix $\Sigma_{\x}$ is
immaterial, in the sense that in the limit as $\epsilon
\rightarrow 0$, then $\widehat{p}\left(s_{1:K}|\y\right)
\rightarrow p\left(s_{1:K}|\y\right)$ assuming sufficient
statistics $\Stats(\x)$. However, in practice, algorithm efficiency is
directly affected by the choice of $\Sigma_{\x}$. We demonstrate this in Section \ref{simulationresults}.

\subsection{Weighting function}
For weighting function $p\left(\y|\x,s_{1:K},h_{1:L},g_{1:L}\right) =
p\left(\y|\x\right)$ we consider the HD weighting function in (\ref{WeightingFunctionHD}) and the SD weighting function
in (\ref{WeightingFunctionSD}).

\subsection{Tolerance schedule}
With a HD weighting function, an MCMC-ABC algorithm can
experience low acceptance probabilities for extended periods,
particularly when the chain explores the tails of the posterior
distribution (this is known as ``sticking'' c.f. \cite{peters2006bim}). In
order to achieve improved chain convergence, we implement an
annealed tolerance schedule during Markov chain burn-in, through
$ \epsilon _{n}=\max \left\{ N-10n,\epsilon^{\min } \right\}$, where $\epsilon_{n}$ is the tolerance at time $n$ in the Markov chain, $N$ is the overall number of Markov chain samples, and $\epsilon^{\min}$ denotes the target tolerance of the
sampler.

There is a trade-off between computational overheads (i.e. Markov
chain acceptance rate) and the accuracy of the ABC posterior
distribution relative to the true posterior. In this paper we
determine $\epsilon^{\min}$ via preliminary analysis of the Markov
chain sampler mixing rates for a transition kernel with
coefficient of variation set to one. In general practitioners will
have a required precision in posterior estimates. This precision
can be directly used to determine, for a given computational
budget, a suitable tolerance $\epsilon^{\min }$.
\subsection{Performance diagnostic}
Given the above mixing issues, one should carefully monitor
convergence diagnostics of the resulting Markov chain for a given
tolerance schedule. For a Markov chain of length $N$, the
performance diagnostic we consider is the autocorrelation
evaluated on $\tilde{N}=N-N_{b}$ post-convergence samples after an
initial burn-in period $N_b$. Denoting by
$\{\theta_i^{[n]}\}_{n=1:\tilde{N}}$ the Markov chain of the
$i$-th parameter after burn-in, we define the autocorrelation
estimate at lag $\tau$ by
\begin{equation}
\widehat{ACF}(\theta_i,\tau) =
\frac{1}{(\tilde{N}-\tau)\hat{\sigma}\left(\theta_i\right)}\sum_{n=1}^{\tilde{N}-\tau}[\theta_i^{[n]}-\widehat{\mu}\left(\theta_i\right)][\theta_i^{[n+\tau]}-\widehat{\mu}\left(\theta_i\right)],
\end{equation}
where $\widehat{\mu}\left(\theta_i\right)$ and
$\hat{\sigma}\left(\theta_i\right)$ are the estimated mean and
standard deviation of $\theta_i$.
\normalsize
\section{Auxiliary variable MCMC approach} \label{AuxiliaryMCMC}
In this section we demonstrate an alternative solution to the previously presented MCMC-ABC detector.
We will show that at the expense of increasing the dimension of the parameter vector, one can develop a standard
MCMC algorithm without the requirement of the ABC methodology.
We augment the parameter vector with the unknown noise
realizations at each relay, $\mathbf{w}_{1:K}$, to obtain a new
parameter vector $\left(\mathbf{s}_{1:K},g_{1:L},h_{1:L},\mathbf{w}_{1:K}\right)$. The resulting posterior distribution
$p\left(\mathbf{s}_{1:K},g_{1:L},h_{1:L},\mathbf{w}_{1:K}|\y\right)$ may then be decomposed into the full conditional
distributions:
\small
\begin{subequations}
\begin{align}
\label{eqn:AuxiliaryGibbs1}
p\left(\mathbf{s}_{1:K}|g_{1:L},h_{1:L},\mathbf{w}_{1:K},\y\right) &\propto
p\left(\y|\mathbf{s}_{1:K},g_{1:L},h_{1:L},\mathbf{w}_{1:K}\right)p\left(\mathbf{s}_{1:K}\right),\\
p\left(g_{1:L}|\mathbf{s}_{1:K},h_{1:L},\mathbf{w}_{1:K},\y\right) &\propto
p\left(\y|\mathbf{s}_{1:K},g_{1:L},h_{1:L},\mathbf{w}_{1:K}\right)p\left(g_{1:L}\right),\\
p\left(h_{1:L}|\mathbf{s}_{1:K},g_{1:L},\mathbf{w}_{1:K},\y\right) &\propto
p\left(\y|\mathbf{s}_{1:K},g_{1:L},h_{1:L},\mathbf{w}_{1:K}\right)p\left(h_{1:L}\right),\\
\label{eqn:AuxiliaryGibbs4}
p\left(\mathbf{w}_{1:K}|\mathbf{s}_{1:K},g_{1:L},h_{1:L},\y\right) &\propto
p\left(\y|\mathbf{s}_{1:K},g_{1:L},h_{1:L},\mathbf{w}_{1:K}\right)p\left(\mathbf{w}_{1:K}\right),
\end{align}
\end{subequations}
\normalsize
which form a block Gibbs sampling framework. Conditioning on the
unknown noise random variables at the relays permits a simple
closed form solution for the likelihood and results in tractable
full conditional posterior distributions for (\ref{eqn:AuxiliaryGibbs1})-(\ref{eqn:AuxiliaryGibbs4}). In this case the likelihood is given by
\begin{equation*}
p\left(\y|\s_{1:K},g_{1:L},h_{1:L},\w_{1:K}\right)
= \prod^L_{l=1}p\left(\y^{(l)}|\s_{1:K},g^{(l)},h^{(l)},\w^{(l)}\right),
\end{equation*}
where
\small
\begin{equation*}
p\left(\y^{(l)}|\s_{1:K},g^{(l)},h^{(l)},\w^{(l)}\right) =
\mbox{\fsc CN}\left(\f^{(l)}\left( \S h^{(l)}+
\W^{(l)}\right)g^{(l)},\sigma^2_v \I\right).
\end{equation*}
\normalsize
The resulting Metropolis-within Gibbs sampler for this block Gibbs
framework is outlined in Algorithm \ref{MCMC-AV algo}, where we
define the joint posterior parameter vector $\Theta =
\left(\S,\G,\H,\W\right)$. The Metropolis-Hastings proposals used
to sample from each full conditional distribution were given by eqs. (\ref{MH_prop1})-(\ref{MH_prop2}), and the additional proposal for the auxiliary variables, given by
\begin{itemize}
	\item Transition kernel for $\W_i$ : draw proposal $\W_i^*$ from distribution 
\begin{align}
\begin{split}
q( \w^* \leftarrow \w_i^{[n-1]})
&= q(l)q(i)q(w_i^{(l)}|\w_i^{[n-1]}) \\
&= \frac{1}{KL} \mbox{\fsc CN}\left(\w_i^{[n-1]},
\sigma^2_{w\_rw}\right).
\end{split}
\end{align}
\end{itemize}
\begin{algorithm}
 \caption{MAP sequence detection algorithm using AV-MCMC}
 				 
 \label{MCMC-AV algo}
     \begin{algorithmic}[1]

\LINE{\textbf{Initialize Markov chain state:}} 
\STATE Initialize
n=1, $\S^{\left[1\right]} \sim p\left(\s\right)$,
$g^{[1]}_{1:L} = \widehat{g}_{1:L}$,
$h^{[1]}_{1:L} = \widehat{h}_{1:L}$,
$\mathbf{W}^{\left(0\right)} \sim p\left(\mathbf{w}\right)$
\FOR{$n=1,\hdots,N$}
\LINE{\textbf{Propose new Markov chain state: $\Theta^{*}$ given $\Theta^{[n-1]}$.}}
\STATE Draw an index $i \sim U\left[1,\ldots,K+2L+KL\right]$ 
\STATE Draw proposal $\Theta^*=\left[\theta^{[n-1]}_{1:i-1},\theta^*,\theta^{[n-1]}_{i+1:K+2L+KL}\right]$
from proposal distribution \\
$q(\theta_i^{[n-1]}
\rightarrow \theta^*)$. \\
(Note, the proposal will depend on which element of the $\Theta$ vector is being sampled.)

\STATE Evaluate the acceptance probability\\
                \begin{equation*}
                \begin{split}
                &\alpha\left(\Theta^{[n-1]},\Theta^*\right) = \min\left\{1,
                \frac{p\left(\bm{\theta}^*|\y \right)q(\bm{\theta}^* \rightarrow \bm{\theta}^{[n-1]})}
                {p\left(\bm{\theta}^{[n-1]}| \y \right) q(\bm{\theta}^{[n-1]} \rightarrow \bm{\theta}^*)}\right\}.
                \end{split}
                \end{equation*}
                \STATE Sample random variate $u$, where $U \sim
U\left[0,1\right]$.

 \IF{ $u \leq \alpha\left(\bm{\Theta}^{[n-1]},\bm{\Theta}^*\right)$}
 \STATE $\bm{\Theta}^{[n]}=\bm{\Theta}^*$ 
 \ELSE
 \STATE $\bm{\Theta}^{[n]}=\bm{\Theta}^{[n-1]}$.
 \ENDIF
\ENDFOR
\end{algorithmic}
\end{algorithm}

The MCMC-AV approach presents an alternative to the
likelihood-free Bayesian model sampler, and produces exact samples
from the true posterior following chain convergence. While the
MCMC-AV sampler still performs joint channel estimation and
detection, the trade-off is that sampling the large number of
extra parameters will typically result the need for longer Markov
chains, to achieve the same performance as the ABC algorithm (in
terms of joint estimation and detection performance). This is
especially true in high dimensional problems, such as when the
sequence of transmitted symbols $K$ is long and the number of
relays $L$ present in the system is large, or when the posterior
distribution of the additional auxiliary variables exhibits strong
dependence.

\section{Alternative MAP detectors and lower bounds} \label{alternative_MAP}
One can define a suboptimal solution to the MAP detector, even
with an intractable likelihood, involving a naive, highly
computational algorithm based on a Zero Forcing (ZF) solution. 
The ZF solution is popular in simple system models where it can be efficient and performs well.

Under a ZF solution one conditions on some knowledge of the
partial channel state information, and then perform an explicit
search over the set of all possible symbol sequences. To our
knowledge a ZF solution for MAP sequence detection in arbitrary
non-linear relay systems has not been defined. Accordingly we
define the ZF solution for MAP sequence detection as the solution
which conditions on the mean of the noise at the relay nodes, and
also uses the noisy channel estimates given by the partial CSI
information, to reduce the detection search space.

\subsection{Sub-optimal exhaustive search Zero Forcing
approach}

In this approach we condition on the mean of the noise
$\W^{(l)}=\textbf{0}$, and use the partial CSI estimates of channels coefficients, $\left\{\widehat{h}^{(l)},\widehat{g}^{(l)}
\right\}_{l=1}^L$, to reduce the dimensionality of the MAP
detector search space to just the symbol space $\Omega$. 
The SES-ZF-MAP sequence detector can be expressed as
\small
\begin{align}
\begin{split}
\widehat{\s} &= \argmax_{\s \in \Omega}\prod^L_{l=1} p\left(\s|\y^{(l)},G^{(l)}=\widehat{g}^{(l)},H^{(l)}=\widehat{h}^{(l)},\W^{(l)}=\textbf{0}\right)\\
&=\argmax_{\s \in \Omega}\prod^L_{l=1} p\left(\y^{(l)}|\s,G^{(l)}=\widehat{g}^{(l)},H^{(l)}=\widehat{h}^{(l)},\W^{(l)}=\textbf{0}\right)
p\left(\s\right).
\end{split}
\end{align}
\normalsize
Thus, the likelihood model results in a complex Gaussian distribution for each relay channel, as follows
\begin{align}
\begin{split}
&p\left(\y^{(l)}|\s,G^{(l)}=\widehat{g}^{(l)},H^{(l)}=\widehat{h}^{(l)},\W^{(l)}=\textbf{0}\right)\\
&=\mbox{\fsc CN}\left(\f^{(l)}\left(\s \widehat{h}^{(l)}\right) \widehat{g}^{(l)},\sigma_v^2 \I\right).
\end{split}
\end{align}
As a result, the MAP detection can be solved exactly using an explicit search.

Note however that this approach to symbol detection also involves a very high computational cost, as one must evaluate the posterior
distribution for all $M^K$ code words in $\Omega$. It is usual for communications systems to utilise $M$ as either $64$-ary PAM or
$128$-ary PAM and the number of symbols can be anything from $K=1$ to $K=20$ depending on the channel capacity budget for the
designed network and the typical operating SNR level. Typically this explicit search is not feasible to perform. However, the
sub-optimal ZF-MAP detector provides a comparison for the MCMC-ABC approach, which at low SNR should be a reasonable upper bound for the SER and for high SNR an approximate optimal solution.

The SES-ZF-MAP sequence detector can be highly sub-optimal for low SNR values. This is trivial to see,
since we are explicitly setting the noise realisations to zero when the variance the noise distribution is large. For the same
reasoning, in high SNR values, the ZF approach becomes close to optimal. 
\subsection{Lower bound MAP detector performance}
We denote the theoretical lower bound for the MAP detector
performance as the oracle MAP detector (OMAP). The OMAP detector
involves conditioning on perfect oracular knowledge of the
channels coefficients $\left\{h^{(l)},g^{(l)}\right\}_{l=1}^L$ and
of the realized noise sequence at each relay $\W^{(l)}$. This
results in the likelihood model for each relay channel being
complex Gaussian, resulting in an explicit solution for the MAP
detector. Accordingly, the OMAP detector provides the lower bound
for the SER performance. 
The OMAP detector is expressed as
\small
\begin{align*}
\begin{split}
\widehat{\s} = \argmax_{\s \in \Omega}\prod^L_{l=1} p\left(\s|\y^{(l)},\mathbf{\Pi}\right)=\argmax_{\s \in \Omega}\prod^L_{l=1} p\left(\y^{(l)}|\s,\mathbf{\Pi}\right)p\left(\s\right),
\end{split}
\end{align*}
\normalsize
where $\mathbf{\Pi} := \left(G^{(l)}=g^{(l)},H^{(l)}=h^{(l)},\W^{(l)}=\W^{(l)}\right)$.\\
In this case, the likelihood model results in a complex Gaussian distribution for each relay channel, as follows
\begin{align}
\begin{split}
&p\left(\y^{(l)}|\s,G^{(l)}=g^{(l)},H^{(l)}=h^{(l)},\W^{(l)}=\W^{(l)}\right)\\
&= \mbox{\fsc CN}\left(\f^{(l)}\left(\s h^{(l)}+\W^{(l)}\right) g^{(l)},\sigma_v^2 \I\right).
\end{split}
\end{align}
However, clearly this is impossible to
evaluate in a real system, since oracular knowledge is not available. 
\section{Results} \label{simulationresults}
We compare the performance of joint channel estimation and
detection under the ABC relay methodology versus the auxiliary
MCMC approach for different model configurations. Additionally, we
compare the detection performance of the ABC relay methodology,
the auxiliary MCMC approach, the optimal oracle MAP sequence
detector and the SES-ZF MAP sequence detector. Each presented
Monte Carlo sampler has a burn-in of $5,000$ iterations followed
by a further $15,000$ recorded iterations ($N=20,000$). Random walk proposal
variances for each of the parameters were tuned off-line to ensure
the average acceptance probability (post burn-in) was in the range
$0.3$ to $0.5$.

The following specifications for the relay system model are used
for all the simulations performed: the symbols are taken from a
constellation which is $4$-PAM at constellation points in
$\{-3,-1,1,3\}$; each sequence contains $K=2$ symbols; the prior
for the sequence of symbols is
$p\left(\left(s_{1}, s_{2}\right) = \left[1,1\right]\right)=p\left(\left(s_{1}, s_{2}\right) =
\left[-1,1\right]\right)=0.3$ otherwise, all other are equi-probable; the partial CSI is given by
$\sigma^2_g = \sigma^2_h =0.1$; the nonlinear relay function is
given by $f\left(\cdot\right)=\tanh\left(\cdot\right)$. These
system parameters were utilised as they allow us to perform the
zero forcing solution, without a prohibitive computational burden.

\subsection{Analysis of mixing and convergence of MCMC-ABC}
We analyze the impact that the ABC tolerance level
$\epsilon^{\min}$ has on estimation performance of channel
coefficients and the mixing properties of the Markov chain for the
MCMC-ABC algorithm. The study involves joint estimation of channel
coefficients and transmitted symbols at an SNR level of $15$dB,
with $L=5$ relays present in the system. We adopt a scaled
Euclidean distance metric with the HD weighting function (\ref{WeightingFunctionHD}) and empirically monitor the mixing of
the Markov chain through the autocorrelation function (ACF).

\begin{figure}
    \centering
        \epsfysize=7cm
        \epsfxsize=9cm
        \epsffile{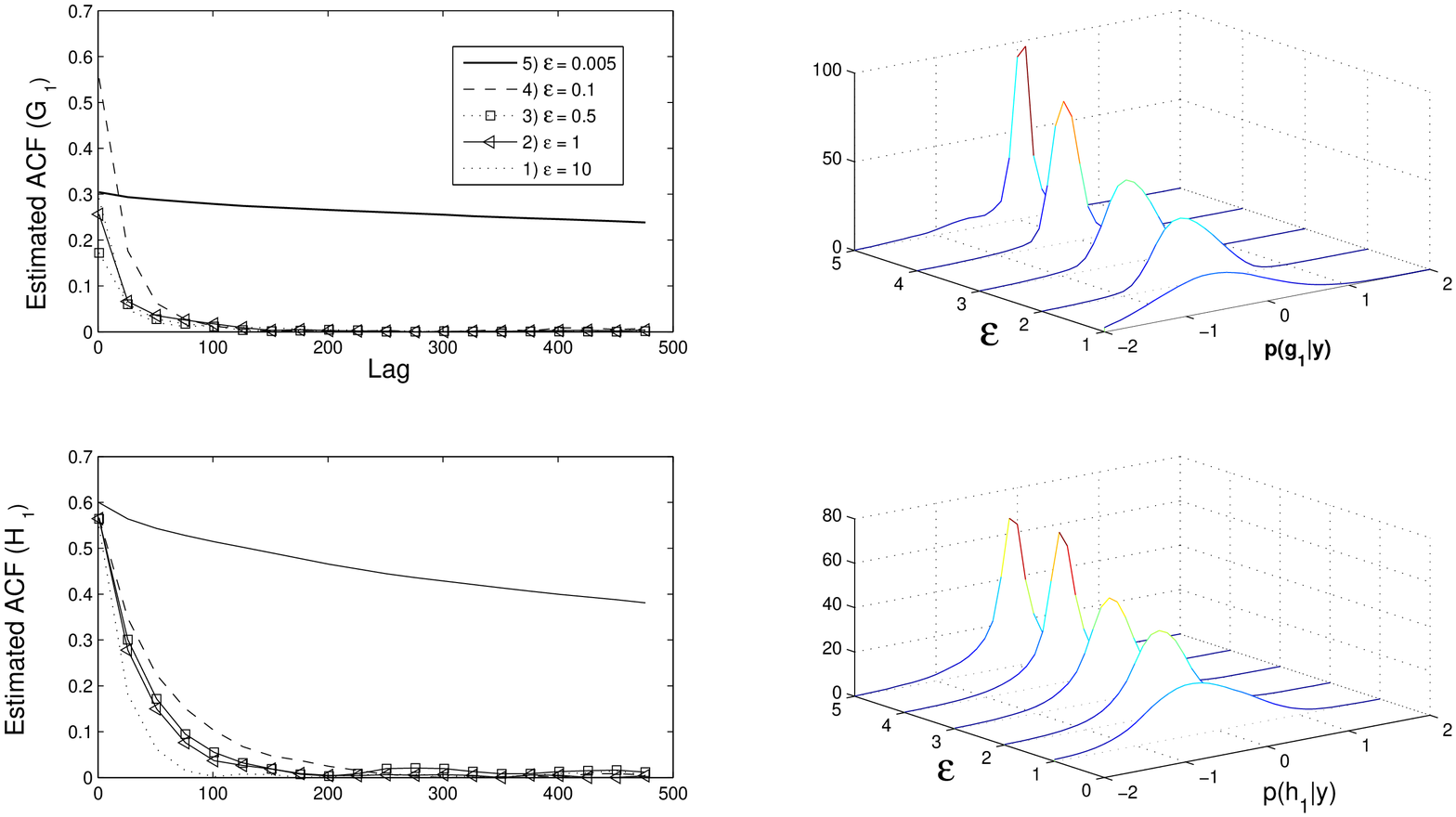}
        \caption{{\small{Comparison of performance for MCMC-ABC with Hard Decision weighting
        and Scaled Euclidean distance metric. Subplots on the left of the image display
        how the estimated ACF changes as a function of tolerance level $\epsilon$ for the
        estimated channels of the relay system. Subplots on the
        right of the image display a sequence of smoothed marginal posterior distribution
        estimates for the first channel of the relay, as the tolerance decreases. Note, the indexing of each marginal
        distribution with labels $1,2,\ldots$ corresponds to the tolerance given in the legend
        on the left hand plots.}}}
    \label{fig:part1}
\end{figure}

In Fig. \ref{fig:part1} we present a study of the ACF of the Markov chains for the channel
estimations of $G_1$ and $H_1$ as a function of the tolerance
$\epsilon$, and the associated estimated marginal posterior
distributions $p(g_1|\Y)$ and $p(h_1|\Y)$.
For large $\epsilon$ the Markov chain mixes over the posterior
support efficiently, since when the tolerance is large, the HD
weighting function and therefore the acceptance probability, will
regularly be non-zero. In addition, with a large tolerance the
posterior almost exactly recovers the prior given by the partial
CSI. This is expected since a large tolerance results in a weak
contribution from the likelihood. As the tolerance $\epsilon$
decreases, the posterior distribution precision increases and
there is a translation from the prior partial CSI channel
estimates to the posterior distribution over the true
generated channel coefficients for the given frame of data
communications.

It is evident that the mixing properties of the MCMC-ABC algorithm
are impacted by the choice of tolerance level. A decreasing
tolerance results in more accurate posterior distribution, albeit
at the price of slower chain mixing. Clearly, the ACF tail decay
rate is significantly slower as the tolerance reduces.

Note, that although the results are not presented here, we also performed
analysis for all aspects of Algorithm 1 under the setting in which
the relay function is linear. We confirmed the MMSE estimates of
the channel coefficients and the MAP sequence detector results
were accurate for a range of SNR values. The results presented
here are for a much more challenging setting in which the relay is
highly non-linear, given by a hyperbolic tangent function.

\subsection{Analysis of ABC model specifications}
We now examine the effect of the distance metric and weighting function on the performance of the MCMC-ABC algorithm as a function of the tolerance.
We consider HD and SD weighting functions with both Mahalanobis and scaled Euclidean distance metrics.
We consider the estimated ACF of the Markov chains of each of the
channel coefficients $G$ and $H$. The SNR level was set to $15$dB and $L=5$ relays were present. 
The results are presented for one channel; since all channels are i.i.d. this will be indicative of the performance of all channels.

For comparison, an equivalent ABC posterior precision should be obtained under each algorithm. Since, the weighting and distance
functions are different, this will result in different $\epsilon^{\min}$ values for each choice in the MCMC-ABC algorithm. As a result, analysis proceeded by first taking a minimum base epsilon value, $\epsilon_{b} = 0.2$ and running the MCMC-ABC with soft decision Gaussian weighting and Mahalanobis distance for $100,000$ simulations. 
We ensured that the average acceptance rate was between $[0.1,0.3]$. This produced a ``true" empirical distribution function (EDF) which we used as our baseline comparison estimate of the true cdf.
Now, for a range of tolerance values $\epsilon_{i}=[0.25,0.5,0.75,1]$ we ran the MCMC-ABC algorithm with soft decision Gaussian weighting and Mahalanobis distance for $20,000$ iterations, ensuring the average acceptance rate was in the interval $[0.1,0.3]$.
This produced a set of random walk standard deviation $\sigma_{b(i)}$, one for each tolerance, that we used for the analysis in the remaining choices of the MCMC-ABC algorithms. For comparison purposes we recorded the estimated maximum error between the estimated EDF for each algorithm and the baseline ``true" EDF. We repeated the simulations for each tolerance on $20$ independent generated data sets.

\begin{figure}
    \centering
        \epsfysize=6cm
        \epsfxsize=8cm
        \epsffile{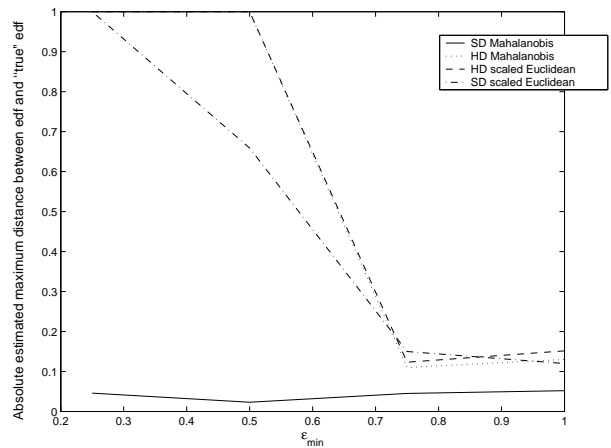}
        \caption{{\small{Maximum distance between the EDF and the baseline ``true" EDF for the first channel, estimated cdf for $G_1$, averaged over 20 independent data realisations.}}}
    \label{fig:ACF}
\end{figure}

In Fig. \ref{fig:ACF} we present the results of this analysis, which demonstrate that the algorithm
producing the most accurate results utilised the soft decision and Mahalanobis distance. The worst performance
involved the hard decision and scaled Euclidean distance. In this case, at low tolerances the average distance between the EDF and the baseline ``true"  EDF was a maximum, since the algorithm was not mixing. This demonstrates that such low
tolerances under this setting of the MCMC-ABC algorithm will
produce poor performance, relative to the soft decision with Mahalanobis distance.
\subsection{Comparisons of detector performance}
Finally, we present an analysis of the symbol error rate (SER)
under the MCMC-ABC algorithm (with a SD weighting function and
Mahalanobis distance), the MCMC-AV detector algorithm, and the
SES-ZF and Oracle detectors. Specifically, we systematically study
the SER as a function of the number of relays, $L\in
\left\{1,2,5,10\right\}$ and the SNR $\in
\left\{0,5,10,15,20,25,30\right\}$.

\begin{figure}
    \centering
        \epsfysize=7cm
        \epsfxsize=9cm
        \epsffile{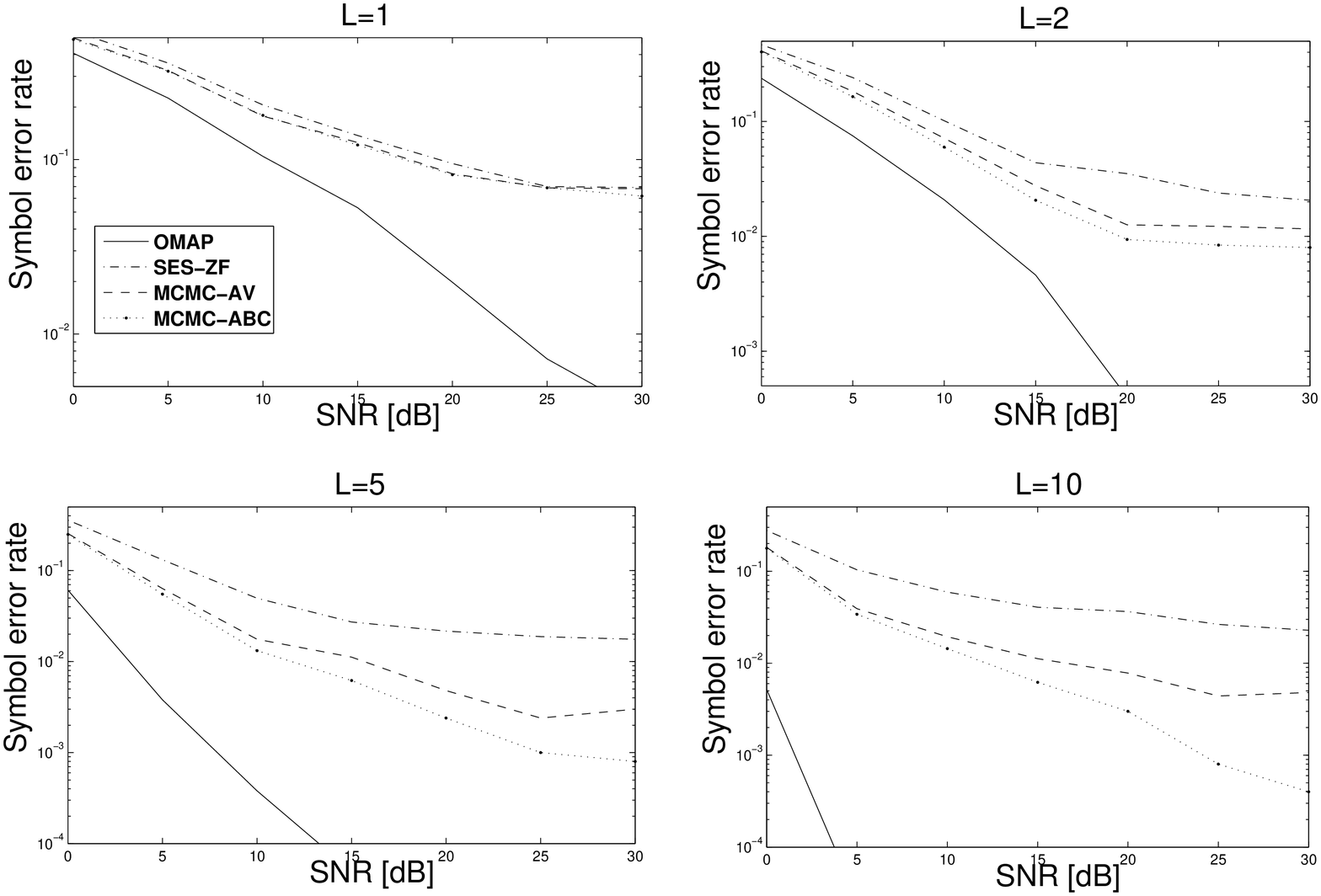}
        \caption{{\small{SER performance of each of the proposed detector schemes as a function of the
        number of relay links, $L$. For each relay set up the SER is reported as a function of the SNR.}}}
    \label{fig:BER1}
\end{figure}

The results of this analysis are presented in Fig. \ref{fig:BER1}.
In summary these results
demonstrate that under our proposed system model and detection
algorithms, spatial diversity stemming from an increasing number
of relays results in measurable improvements in the SER
performance. For example Fig. \ref{fig:BER1} demonstrates that
for $L=1$, there is an insignificant difference between the
results obtained for algorithms MCMC-ABC, MCMC-AV and SES-ZF.
However, as $L$ increases, SES-ZF has the worst performance and
degrades relative to the MCMC-based approaches, demonstrating the
utility obtained by developing a more sophisticated detector
algorithm. It is clear that the SES-ZF suffers from an error-floor
effect: as the SNR increases the SER is almost constant for SNR
values above $15$dB.

Also in Fig. \ref{fig:BER1} the two MCMC-based approaches
demonstrate comparable performance for small $L$. However as $L$
increases, in the high SNR region, the difference in performance
between the MCMC-AV and MCMC-ABC algorithms increases. 
This could be due to the greater numbers of auxiliary parameters to be estimated in the auxiliary-based approach as $L$
increases. In particular we note that adding an additional relay introduces $K$ additional nuisance parameters into the auxiliary
model posterior.

\section{Conclusions} \label{conclusions}
In this paper, we proposed a novel cooperative relay system model
and then obtained novel detector algorithms. In particular, this
involved an approximated-MAP sequence detector for a coherent
multiple relay system, where the relay processing function is
non-linear. Using the ABC methodology we were able to perform
"likelihood free" inference on the parameters of interest.
Simulation results validate the effectiveness of the proposed
scheme. In addition to the ABC approach, we developed an
alternative exact novel algorithm, MCMC-AV, based on auxiliary
variables. Finally, we developed a sub-optimal zero forcing
solution. We then studied the performance of each algorithm under
different settings of our relay system model, including the size
of the network and the noise level present. 
As a result of our findings, we recommend the use of the MCMC-ABC detector especially when there are many relays present at the network, or the number of symbols transmitted in each frame is large. In settings where the number of relays is moderate, one could consider using the MCMC-AV algorithm, as its performance was on par with the MCMC-ABC results and does not involve an ABC approximation.

Future research includes the design of detection algorithms for
relay systems with partial CSI in which the relay system topology
may contain multiple hops on a given channel, or the relay network
topology may be unknown. This can include aspects such as an
unknown number of relay channels or hops per relay channel.

\vspace{0.1cm} \noindent \textbf{{\large
{Acknowledgements}}}\newline \noindent 
The authors would also like to thank the reviewers for their thorough and thoughtful comments.
GWP thanks Prof. Arnaud
Doucet and Dr. Mark Briers for useful discussion and the
Department of Mathematics and Statistics at UNSW (through an APA)
and CMIS, CSIRO for financial support. This material was based
upon work partially supported by the National Science Foundation
under Grant DMS-0635449 to the Statistical and Applied
Mathematical Sciences Institute, North Carolina, USA. Any
opinions, findings, and conclusions or recommendations expressed
in this material are those of the author(s) and do not necessarily
reflect the views of the NSF. SAS and YF were supported by the
Australian Research Council through the Discovery Project scheme (DP0664970 and DP1092805).
IN and JY were supported by the Australian Research Council through the Discovery Project scheme (DP0667030).

\bibliographystyle{IEEEtran}
\bibliography{reference_all}
\end{document}